\documentclass{LMCS}

\usepackage{enumerate}

\newcommand\Set{\mathbf{Set}}
\newcommand\Rel{\mathbf{Rel}}
\newcommand\Coh{\mathbf{Coh}}
\newcommand\Fin{\mathbf{Fin}}
\newcommand\FinCoh{\mathbf{FinCoh}}
\newcommand\N{\mathbf{N}}
\newcommand\R{\mathbf{R}}
\newcommand\B{\mathbf{B}}

\newcommand\SIGMA{\mathsf{\Sigma}}
\newcommand\PI{\mathsf{\Pi}}
\newcommand\BLANK{\texttt{\char"5F}}
\newcommand\eqdef{\stackrel{\smash{\text{\tiny\sf def}}}{=}}
\newcommand\subinfty{\mathrel{\smash{\lower4pt\hbox{$\scriptscriptstyle\infty$}}%
                      \mskip-14mu%
                       \raise0pt\hbox{$\subset$}}}

\newcommand\ssup{{$\scriptstyle(\supseteq)$}}
\newcommand\ssub{{$\scriptstyle(\subseteq)$}}

\newcommand\Tensor{\otimes}
\newcommand\Plus{\oplus}
\newcommand\bbot{{\bot\mskip-12mu\bot}}

\DeclareMathOperator\Cl{\mathcal{C}}
\DeclareMathOperator\FC{\mathcal{F}}
\DeclareMathOperator\Pow{\mathcal{P}}
\DeclareMathOperator\Mf{\mathcal{M}_{\mskip-2.5mu\mathit{f}}}

\def\doi{7 (3:15) 2011}
\lmcsheading%
{\doi}
{1--7}
{}
{}
{Feb.~\phantom03, 2011}
{Sep.~22, 2011}
{}   

\begin{document}

\title{From Coherent to Finiteness Spaces}

\author[P.~Hyvernat]{Pierre Hyvernat}
\address{Laboratoire de Math\'ematiques, Universit\'e de Savoie, 73376 Le Bourget-du-Lac Cedex, France.}
\email{Pierre.Hyvernat@univ-savoie.fr}
\thanks{This work has been partially funded by the French project choco (\texttt{ANR-07-BLAN-0324}).}


\begin{abstract}
This short note presents a new relation between coherent spaces and finiteness
spaces. This takes the form of a functor from~$\Coh$ to~$\Fin$ commuting with
the additive and multiplicative structure of linear logic.  What makes this
correspondence possible and conceptually interesting is the use of the
infinite Ramsey theorem. Along the way, the question of the cardinality of the
collection of finiteness spaces on $\N$ is answered.

\smallbreak
\noindent
Basic knowledge about coherent spaces and finiteness spaces is assumed.
\end{abstract}

\maketitle

\section*{Introduction}

The category of coherent spaces was the first denotational model for linear
logic~\cite{Girard87}: the basic objects are countable reflexive non-oriented
graphs, and we are more specifically interested by their \emph{cliques}
(complete subgraphs). If~$C$ is such a graph, we write~$\Cl(C)$ for the
collection of its cliques.
Coherent spaces enjoy a rich algebraic structure where the important
operations are:
\begin{iteMize}{$\bullet$}
  \item the (reflexive closure of the) complement, written~$C_1^\bot$;
  \item the product, written~$C_1\Tensor C_2$;
  \item the disjoint union, written~$C_1\Plus C_2$.
\end{iteMize}
If one forgets about edges and only looks at vertices, the corresponding
operations are simply the identity, the usual cartesian product~``$\times$''
and the disjoint union~``$\Plus$''.

\medbreak
More recently, T. Ehrhard introduced the notion of finiteness
space~\cite{Ehrhard05} to give a model for the differential
$\lambda$-calculus~\cite{EhrhardRegnier03}, which can be seen as an enrichment
of linear logic. The point that interests us most here is that the collection
of \emph{finitary sets} of a finiteness space is closed under finite unions.
This is definitely not the case with the cliques of a coherent space.
This property is crucial for the interpretation of non-deterministic sums of
terms~\cite{Ehrhard05,Vaux09}, which correspond in the models to linear
combinations of simple terms. In this note, we only look at a qualitative
version, where coefficients play no role. In other words, coefficients live in
the \emph{rig} (ring without \emph{n}egatives)~$\{0,1\}$ with~$1+1=1$.

Very briefly, a finiteness space is given by a countable set~$|\mathcal{F}|$,
called its \emph{web}, and a collection~$\mathcal{F}$ of subsets
of~$|\mathcal{F}|$, called the \emph{finitary sets}, which satisfies
\[
  \mathcal{F}^{\bbot\bbot}
  \quad=\quad
  \mathcal{F}
\]
with
\[
  \mathcal{D}^\bbot
  \quad\eqdef\quad
  \bigl\{x\ \bigm|\ \forall y\in\mathcal{D}, \#(x\cap y)<\aleph_0\bigr\}
\]
where~$\#(x)$ is the cardinal of~$x$, and~$\aleph_0$ is the least infinite
cardinal. In natural language, the crucial property is that
whenever~$x\in\mathcal{F}$ and~$y\in\mathcal{F}^\bbot$, the
intersection~$x\cap y$ is finite.
Algebraic constructions similar to the ones for coherent spaces can be defined
on finiteness spaces, and they are characterized by:
\begin{iteMize}{$\bullet$}
  \item the dual $\mathcal{F}^{\bbot}$;

  \item the tensor $\mathcal{F}_1\Tensor\mathcal{F}_2 \eqdef \{r\ |\ \pi_1(r)\in
  \mathcal{F}_1,\ \pi_2(r)\in \mathcal{F}_2\}$;

  \item the coproduct $\mathcal{F}_1\Plus\mathcal{F}_2 \eqdef \{x_1\uplus x_2\ |\
  x_1\in\mathcal{F}_1,\ x_2\in\mathcal{F}_2\}$.
\end{iteMize}
Here again, if one forgets about finitary sets and only looks at the webs of
finiteness spaces, the corresponding operations are just the identity, the
usual cartesian product and the disjoint union.

\subsubsection*{Remarks}

\begin{enumerate}[(1)]

  \item The operations on finiteness spaces are actually defined in a
    way that makes it clear that they yield finiteness spaces. They are then
    proved to be equivalent to the definitions given above~\cite{Ehrhard05}.

  \smallbreak
  \item\label{rk:closure} Any operator of the form $\mathcal{X} \mapsto
    \mathcal{X}^\ast = \{y\ |\ \forall x\in \mathcal{X},\ R(x,y)\}$ is
    contravariant with respect to inclusion and yields a \emph{closure
    operator} when applied twice.
    In particular, any set of the form~$\mathcal{Y}=\mathcal{X}^\ast$
    satisfies~$\mathcal{Y}=\mathcal{Y}^{\ast\ast}$.

  \smallbreak
  \item There is another presentation of coherent spaces that closely
    matches the definition of finiteness spaces: a coherent space is given by
    a collection~$\mathcal{C}$ of subsets of~$|\mathcal{C}|$ which
    satisfy~$\mathcal{C}^{\ast\ast}=C$, where~$\mathcal{D}^\ast = \{x\ |\
    \forall y\in\mathcal{D}, \#(x\cap y)\leq1\}$.

\end{enumerate}


\section{From Coherence to Finiteness}

The idea is rather simple: we would like to close the collection of cliques of
a coherent space under finite unions. Unfortunately (but unsurprisingly), the
notion of ``finite unions of cliques'' is not very well behaved, especially
with respect to the dual. Recall that an anticlique of~$C$ (also called
independent, or stable sets) is a clique in~$C^\bot$. We consider the
following notion:

\begin{defi}
  If $C$ is a coherent space, we call a subset of $|C|$ \emph{finitely
  incoherent} if it doesn't contain infinite anticliques. We write $\FC(C)$
  for the collection of all finitely incoherent subsets of $C$.
\end{defi}

\noindent
The next lemma follows directly from the definition.
\begin{lem}\label{l:easy}\leavevmode
  \begin{enumerate}[\em(1)]
    \item Any finite subset of $|C|$ is finitely incoherent;
    \item a subset of a finitely incoherent subset is finitely incoherent;
    \item finitely incoherent subsets are closed under finite unions;
    \item\label{l:easy_inclusion} any clique is finitely incoherent.
  \end{enumerate}
\end{lem}

\noindent
Note however that a finitely incoherent set needs not be a finite union of
cliques: take for example the graph composed of the disjoint union of all the
complete graphs $K_n$ for $n\geq1$. This graph doesn't contain an infinite
clique, but it is not a finite union of anticliques; so, its dual is finitely
incoherent but is not a finite union of cliques.

\medbreak
The next lemma is more interesting as it implies that the collection of
finitely incoherent subsets forms a finiteness space. 

\begin{lem}\label{l:neg_aux}
  If $C$ is a coherent space, we have:
  \[
    \Cl(C)^\bbot \quad=\quad \FC(C^\bot)
  \  .
  \]
\end{lem}

\proof\leavevmode

\begin{iteMize}{\ssub}

  \item Let $x$ be in $\Cl(C)^\bbot$, and suppose, by contradiction, that
  $x$ is not in $\FC(C^\bot)$, i.e., $x$ contains an infinite anticlique $y$ of
  $C^\bot$. This set $y$ is a clique in $C$, i.e., $y\in \Cl(C)$.
  Since~$x\cap y=y$ is infinite, this contradicts the hypothesis that
  $x\in\Cl(C)^\bbot$.

  \medbreak

  \item[\ssup] Let $x$ be finitely incoherent in~$C^\bot$, i.e., $x$ doesn't
  contain an infinite clique of~$C$; let~$y$ be in~$\Cl(C)$. Since~$x\cap y
  \in \Cl(C)$ and~$x\cap y$ is contained in $x$, it cannot be infinite. This
  shows that~$x\in \Cl(C)^\bbot$.

\end{iteMize}

\qed

\noindent
By remark~\ref{rk:closure} on page~\pageref{rk:closure}, we thus get the expected corollary:
\begin{cor}
  If $C$ is a coherent space, then $\FC(C)$ is a finiteness space.
\end{cor}

\noindent
What was unexpected is the following:
\begin{lem}\label{l:neg}
  If $C$ is a coherent space, then:
  \[
    \FC(C^\bot)
    \quad=\quad
    \FC(C)^\bbot
    \ \hbox{.}
  \]
\end{lem}

\proof

Because of the previous lemma, and because $\BLANK^\bbot$ is contravariant with
respect to inclusion, we only need to show that
$\Cl(C)^\bbot\subseteq\FC(C)^\bbot$. Suppose that $x\in \Cl(C)^\bbot$, and
let~$y\in\FC(C)$; we need to show that $x\cap y$ is finite.

\begin{iteMize}{$\bullet$}
  \item Since $x\cap y\subseteq y\in\FC(C)$, $x\cap y$ cannot contain an
  infinite anticlique;

  \item since $x\cap y\subseteq x\in \Cl(C)^\bbot$, $x\cap y$ cannot contain
  an infinite clique.
\end{iteMize}
Those two points imply, by the infinite Ramsey
theorem,\footnote{\emph{Infinite Ramsey theorem:} suppose~$G$ is a countably
infinite set, then, for every assignment of~$c$ colors to the subsets of~$G$
of cardinality~$n$, there is an infinite $I\subseteq G$ s.t. all subsets
of~$I$ of cardinality~$n$ have the same color. (Refer to~\cite{Ramsey30} or
one of the many textbooks on combinatorics covering it.) For~$n=2$ and~$c=2$,
it amounts to ``each countably infinite graph has an infinite clique or an
infinite anticlique''.} that $x\cap y$ is finite.

\qed

\medbreak
The other linear connectives are similarly behaved with respect to the notion
of finitely incoherent sets. We have:
\begin{lem}\label{l:tensor}
  If $C_1$ and $C_2$ are coherent spaces, then we have both
  \[
    \FC(C_1\Plus C_2)
    \quad=\quad
    \FC(C_1)\Plus\FC(C_2)
    \ \hbox{,}
  \]
  and
  \[
    \FC(C_1\Tensor C_2)
    \quad=\quad \FC(C_1)\Tensor\FC(C_2)
  \]
  where the connectives on the left are the coherent spaces' ones, and the
  connectives on the right are the finiteness spaces' ones.
\end{lem}

\proof

The $\Plus$ part is direct; for the $\Tensor$ part, recall that~$r\in
\FC(C_1)\Tensor\FC(C_2)$ is equivalent to~$\pi_1(r)\in\FC(C_1)$
and~$\pi_2(r)\in\FC(C_2)$.

\begin{iteMize}{\ssub}

  \item Suppose $r$ doesn't contain an infinite anticlique;
  neither~$\pi_1(r)$ nor~$\pi_2(r)$ can contain an infinite anticlique, as it
  would imply the existence of an infinite anticlique in~$r$.

  \smallbreak
  \item[\ssup] Suppose that~$r\in\FC(C_1)\Tensor\FC(C_2)$ contains an infinite
  anticlique~$r'$ of~$C_1\Tensor C_2$.  At least one of~$\pi_1(r')$
  or~$\pi_2(r')$ must be infinite, otherwise, $r'$ itself would be finite.
  Suppose~$\pi_1(r')$ is infinite; because~$\pi_1(r')\subseteq\pi_1(r)$, it
  cannot contain an infinite anticlique. By the infinite Ramsey theorem, it
  thus contains an infinite clique $x$. For each~$a\in x$, chose one
  element~$b$ inside the fiber~$r'(a)=\{b\ |\ (a,b)\in r'\}$. Two
  such~$b$'s cannot be coherent as it would contradict the fact that~$r'$ is an
  anticlique. In particular, all such~$b$'s are distinct. We have constructed
  an infinite anticlique in~$\pi_2(r')\subseteq\pi_2(r)\in\FC(C_2)$.
  Contradiction!

\end{iteMize}

\qed

\noindent
For finiteness spaces, the operation~$\Plus$ coincides with its dual
\cite{Ehrhard05}. In particular, for finiteness spaces coming from from
coherent spaces, we have
\[
  \big(\FC(C_1)\Plus\FC(C_2)\big)^\bbot
  \quad=\quad
  \FC(C_1)^\bbot\Plus\FC(C_2)^\bbot
  \ \hbox{.}
\]
Lemmas~\ref{l:neg} and~\ref{l:tensor} thus imply that
\[
  \FC\big((C_1\Plus C_2)^\bot\big)
  \quad = \quad
  \FC\big(C_1^\bot\Plus C_2^\bot\big)
  \ \hbox{.}
\]
The direct proof of this equality is also quite easy.

\bigbreak\noindent
Both coherent spaces and finiteness spaces form categories, where:
\begin{iteMize}{$\bullet$}
  \item a morphism from $C$ to $D$ in~$\Coh$ is a clique in~$(C\Tensor
  D^\bot)^\bot$,

  \item a morphism from $\mathcal{F}$ to $\mathcal{G}$ in~$\Fin$ is a finitary set
  in~$(\mathcal{F}\Tensor \mathcal{G}^\bbot)^\bbot$.
\end{iteMize}
In both cases, morphisms are special relations between webs and
composition is the usual composition of relations:
\[
  r \circ s
  \quad\eqdef\quad
  \bigl\{(a,c) \ \bigm|\ \exists b\,(a,b)\in s \text{ and } (b,c)\in r \bigr\}
  \ \hbox{.}
\]
From all the above, we can
conclude that:
\begin{prop}
  The operation~$\FC(\BLANK)$ can be lifted to a functor from $\Coh$ to $\Fin$:
  \begin{enumerate}[\em(1)]
    \item it sends $C$ to $\FC(C)$
    \item and $r\in\Coh[C,D]$ to $r\in\Fin[\FC(C),\FC(D)]$.
  \end{enumerate}
  Moreover, this functor commutes with
  $\BLANK^\bot$,~$\BLANK\Tensor\BLANK$ and~$\BLANK\Plus\BLANK$.
\end{prop}
This functor is faithful (but not full), and it is not injective on objects as
adding or removing any finite number of edges to a coherent space doesn't
change its image via~$\FC(\BLANK)$. Moreover, this functor commutes with the
forgetful functors from~$\Coh$ and~$\Fin$ to~$\Rel$, the category of sets and
relations.

\medbreak
In a sense, coherent spaces allow one to define a collection of simple
finiteness spaces. An informal argument regarding this simplicity can be
found in the following remark: the logical complexity of the formula
expressing~``$x\in \mathcal{A}^{\bbot\bbot}$'', i.e., ``$x$ is finitary
with respect to~$\mathcal{A}$'' changes when~$\mathcal{A}$ comes from a
coherent space. If we write~$y\subinfty x$ for ``$y$ is an infinite subset
of~$x$'', we have~\cite{Ehrhard05}:
\[
  x \in \mathcal{A}^{\bbot}
  \qquad \iff \qquad
  \forall y\subinfty x\quad%
  y\notin \mathcal{A}
\]
whenever~$\mathcal{A}$ is downward closed. Thus:
\[
  x \in \mathcal{A}^{\bbot\bbot}
  \qquad \iff \qquad
  \forall y\subinfty x\quad%
  \exists z\subinfty y\quad%
  z\in \mathcal{A}
  \ \hbox{.}
\]
Because $y\subinfty x$ is a~$\SIGMA^1_1$-formula (no universal second-order
quantifiers), ``$x\in\mathcal{A}^{\bbot\bbot}$'' is a $\PI^1_2$-formula
(second-order quantifiers are~$\forall\exists$). Note that even
if~$\mathcal{A}$ isn't downward closed, the formula
expressing~``$x\in\mathcal{A}^{\bbot\bbot}$'' is still a $\PI^1_2$-formula.
For the particular case when $\mathcal{A}$ is the set of cliques of a
coherent spaces~$C$, we obtain
\[
  x \in \mathcal{A}^{\bbot\bbot}
  \qquad \iff \qquad
  x \in \Cl(C^\bot)^\bbot
  \qquad \iff \qquad
  \forall y\subinfty x \quad %
  \exists\, a,b \in y \quad (a,b) \in C
\]
which is only a $\PI^1_1$-formula.


\section{Cardinality of Finiteness Spaces}

So, coherent spaces can be used to define a collection of finiteness spaces
closed under the linear operations ($\BLANK^\bot$, $\BLANK\Tensor\BLANK$ and
$\BLANK\Plus\BLANK$).  It is natural to ask whether all finiteness spaces can be
obtained in this way.  The previous informal remark about the logical
complexity of coherence versus finiteness points toward a negative answer.
Here is a formal proof which also answers a question raised by T. Ehrhard:
\begin{prop}
  If $A$ is infinite countable, the cardinality of finiteness spaces on~$A$ is
  exactly that of~$\Pow(\Pow(A))$. The cardinality doesn't change if we
  consider finiteness spaces up-to isomorphisms, i.e., up-to permutations of~$A$.
\end{prop}

\noindent
Since the cardinality of coherent spaces on~$A$ is the same as that
of~$\Pow(A\times A)\simeq\Pow(A)$, we can conclude that:

\begin{cor}
  If $A$ is infinite countable, there are strictly more finiteness spaces on
  $A$ than coherent spaces on $A$.
\end{cor}

\proof
Let~$A$ be infinite countable; up-to isomorphism, we can assume that~$A=
\B^{<\omega}$, the set of finite sequences of bits. If~$x$ is an
\emph{infinite} sequence of bits, write~$x^\downarrow$ for the set of finite
approximations of~$x$; and if~$X$ is a set of such ``real numbers'',
write~$X^\downarrow$ for the set~$\{ x^\downarrow\ |\ x\in X\}$. We
have~$X^\downarrow\subset \Pow(A)$ for any such set~$X$.

\smallbreak
Suppose now that $X\neq X'$ with, for example, $x\in X$ but $x\notin X'$.
Since~$x^\downarrow$ is infinite and~$x^\downarrow \in X^\downarrow$, we have
$x^\downarrow\notin X^{\downarrow\bbot}$. However since two different reals
must differ on some finite approximation, we have that~$x^\downarrow\in
X'^{\downarrow\bbot}$. Thus, the finiteness spaces~$(A,X^{\downarrow\bbot})$
and~$(A,X'^{\downarrow\bbot})$ differ.

This defines an injective map~$X\mapsto (A,X^{\downarrow\bbot})$ from
arbitrary sets of reals to finiteness spaces on~$A$. This shows that
finiteness spaces on $A$ have at least the same cardinality
as~$\Pow(\R)\simeq\Pow(\Pow(A))$. Since it cannot be more than that, we have
equality.

\smallbreak
An isomorphism in the category~$\Fin$ is a particular relation with a left
and right inverse. This implies that it is in fact the graph of a bijection,
and thus, two finiteness spaces~$\mathcal{F}_1$ and~$\mathcal{F}_2$ on a
set~$A$ are isomorphic if and only if there is a bijection~$\sigma : A\to A$
such that
\[
  \mathcal{F}_2
  \quad=\quad
  \sigma \cdot \mathcal{F}_1
  \eqdef
  \bigl\{ \sigma(x)\ \bigm|\ x\in\mathcal{F}_1\bigr\}
\]
where~$\sigma(x)$ when~$x\subseteq A$ is simply the direct image of the
set~$x$. Because the cardinality of each equivalence class is at
most~$\#\big(\Pow(A)\big)$ (this is the cardinality of permutations on~$A$),
and since~$\kappa\times\#\big(\Pow(A)\big)=\max\big(\kappa,\#(\Pow(a))\big)$,
there must be at least~$\#\big(\Pow(\Pow(A))\big)$ such equivalence classes to
cover the whole collection of finiteness spaces.

The cardinality of finiteness spaces on~$A$ up-to isomorphism is thus the
same as that of finiteness spaces on~$A$ up-to plain
equality:~$\#\big(\Pow(\Pow(A))\big)$.

\qed
It is slightly interesting to note that the same reasoning doesn't apply to
higher cardinalities since $\#(A^{<\omega})=\#(A)$ if $A$ is uncountable.


\section*{Conclusion}

The situation with respect to full linear logic isn't totally clear.  We have
a base category~$\FinCoh$ with:
\begin{iteMize}{$\bullet$}
  \item coherent spaces as objects
  \item and finitely incoherent linear maps as morphisms:
  $\FinCoh[C,D]=\FC\big((C\Tensor D^\bot)^\bot\big)$.
\end{iteMize}

\noindent
This category is a linear, full subcategory of the category~$\Fin$ of
finiteness spaces.

\medbreak
Lifting the usual set-based notion of exponentials for coherent spaces to this
category is impossible:  because the web of~$!C$ is the collection of finite
cliques of~$C$ (uniformity), this construction isn't even functorial. Take for
example~$K_n$ and~$K_n^\bot$: since their sets of vertices have the same
finite cardinality, they are isomorphic in~$\FinCoh$. However~$!K_n$
and~$!(K_n^\bot)$ have sets of vertices of different cardinalities,
namely~$2^n$ and~$n+1$: they cannot be isomorphic in~$\FinCoh$.\footnote{An
isomorphism in~$\FinCoh$ is in particular an isomorphism in~$\Rel$ which is an
isomorphism in~$\Set$.}

\medbreak
The multiset-based notion of exponentials, where one defines the web~$|!C|$ to
be the collection of finite multisets whose support is a clique, doesn't seem
to help. It is not functorial in any canonical way, as shown by the same
example of $K_n \simeq K_n^\bot$: the corresponding sets of vertices
for~$!K_n$ and~$K_n^\bot$ are~$\Mf\{1,\dots,n\}$
and~$\Mf\{1\}\cup\cdots\cup\Mf\{n\}$. Note that in both cases, the
collection of finitely incoherent sets consists of \emph{all} the subsets of
the web, so that a non-canonical isomorphism is still possible. However, if
such an isomorphism exists, it doesn't commute to the forgetful functors to the
category of sets and relations.

\medbreak Using the non-uniform coherent
spaces~\cite{BucciarelliEhrhard01} isn't a solution either as it
introduces a third relation: neutrality. A non-uniform coherent space
is thus a non-oriented graph with two kinds of edges: strict coherence
edges and neutral edges. Neutral edges are left unchanged by duality
and we take the complement of the rest. In the usual coherent spaces,
the only neutral edges are the loops around vertices. The natural
notion is to define a clique as a set of vertices that are pairwise
coherent or neutral, but one could also consider ``strict''
cliques. Note however that in this case, a singleton cannot be both a
clique and an anticlique as in usual coherent spaces.  Thus, there are
two possible definitions of~$\Cl(C)$:
\[
  \Cl(C)
  \quad\eqdef\quad
  \begin{cases}
    \text{cliques (pairwise coherence or neutrality)}\\
    \text{strict cliques (pairwise strict coherence)}
  \end{cases}
  \ \hbox{,}
\]
and similarly, there are two possible definitions of~$\FC(C)$:
\[
  \FC(C)
  \quad\eqdef\quad
  \begin{cases}
    \text{$x$'s that do not contain anticliques}\\
    \text{$x$'s that do not contain strict anticliques}
  \end{cases}
  \ \hbox{.}
\]
None of the four possibilities enjoys the adequate properties.
When using cliques and anticliques, point~\ref{l:easy_inclusion}
of Lemma~\ref{l:easy} fails: we do not have~$\Cl(C)\subseteq\FC(C)$.
The cliques/strict anticliques and strict cliques/anticliques versions fail
at Lemma~\ref{l:neg_aux}: one inclusion or the other doesn't hold.
With the strict versions of~$\Cl(C)$ and~$\FC(C)$, we go as far as the proof of
Lemma~\ref{l:neg}. However, 
\begin{iteMize}{$\bullet$}
  \item $x\cap y$ doesn't contain an infinite strict anticlique
  \item and~$x\cap y$ doesn't contain an infinite strict clique
\end{iteMize}
only implies, by the infinite Ramsey theorem for three colors, that~$x\cap y$
is finite \emph{or contains an infinite set of pairwise neutral vertices.}

\medbreak

Finding an appropriate notion of exponential to extend this category to a
model of the algebraic $\lambda$-calculus, or better yet, of the differential
$\lambda$-calculus is thus left open at this point.


\bibliographystyle{amsplain}
\bibliography{lmcs-coherence}

\end{document}